\documentclass[12]{article}

\usepackage{amsfonts}
\usepackage{amssymb}
\usepackage{amsmath}
\usepackage{epsfig,multicol}
\usepackage{bbold}
\usepackage{pstricks}

\title{$\mathcal{PT}$-symmetry in quasi-integrable models}

\author{P. E. G. Assis \\
Instituto de F\'{i}sica e Qu\'imica \\
Universidade Federal de Goi\'as, \\
Catal\~ao - Goi\'as - Brazil\\
E-mail: paulo.assis@ufg.br}

\date{}

\begin{document}

\setcounter{equation}{0}

\maketitle

\begin{abstract}
We reinforce the observations of almost stable scattering in nonintegrable models and show that $\mathcal{PT}$-symmetry can be used as a guiding principle to select relevant systems also when it comes to integrability properties.
We show that the presence of unbroken $\mathcal{PT}$-symmetry in classical field theories produces quasi-integrable excitations with asymptotically conserved charges.

\end{abstract}

\section{Introduction}\label{secintro}

\qquad Since it was first conjectured \cite{BenderBoettcher}, the use of $\mathcal{PT}$-symmetry as a guiding principle to identify potentially relevant physical systems has been applied to various quantum settings but there have been also investigations on the effect such a symmetry has at the classical level.

A vast range of classical models was studied, from point particle systems \cite{Hook} to classical field theories \cite{BenderKdV,FringKdV,AFKdV, BenderCooper, AFCompactons}.
An interesting observation was that the requirement of symmetry under parity and time reversal of Calogero-like particles would imply that a class of solitons of the Boussinesq equation would have real energies \cite{AFCalogero}.

At that time it was noticed that most classical field theories of interest were described by equations which were $\mathcal{PT}$-symmetric.
There have been many relevant works investigating $\mathcal{PT}$-symmetric integrable models, e.g. \cite{BagchiFringKdV, CavaFringBagchi, CavagliaFring, PEGA21, PTdimer, NonlDynPT, KinkBreathPT,NonlWang, DisSolVortPT, PTBreath}.
However, this property has not been scrutinised thoroughly enough and there are aspects which remain little explored.
For instance the transition between the broken and the broken phases  of the $\mathcal{PT}$-symmetry has not been fully understood.

In a quantum system, it was shown \cite{BenderBoettcher} that  the invariance of the Hamiltonian with respect to parity and time reversal is not enough to guarantee the reality of energies.
In order to have real eigenvalues it is necessary to have invariant eigenstates as well. If this is the case, eigenenergies are real, otherwise they would come in complex conjugate pairs.

Classically, what are the effects of being in the unbroken or broken phases of a $\mathcal{PT}$-symmetric model?
It turns out this symmetry can impose restrictions on the phase space which can mimic those of integrable systems giving rise to almost stable solitonic solutions.
Quasi-integrable models have been studied independently, as in \cite{FerrWojZak}, since a particular generalisation of the sine-Gordon model was proposed \cite{Bazeia}.
Initially just a numerical observation, the appearance of quasi-integrable behaviour has been more carefully investigated and a deeper comprehension of the reasons why charges were almost conserved was pursued.
This led to an interesting algebraic description of such a phenomenon in \cite{FerrWojZak}, where  the authors identified the intriguing role played by $\mathcal{PT}$.

Here we show how one can understand the quasi-integrability of unbroken $\mathcal{PT}$-symmetric problems in terms of the flat gauge connections in the Lax formulation of intgrable models. In the next section, we motivate the present investigation by showcasing a series of well known integrable nonlinear partial differential equations with a few important solutions and call attention to the space-time symmetries of those.
Then, in section \ref{Cons} we review the construction of conserved charges based on a path-independent solution of Lax's auxiliary problem.
These tools are used in section \ref{BSG}, where a particular nonintegrable model is investigated and shown to possess almost $\mathcal{PT}$-symmetric excitations which lead to the quasi-conservation of integrals of motion. In section \ref{Spec} we construct the approximate conserved charges based on the asymptotic validity of the inverse scattering method and finally, in section \ref{numsec}, a numerical procedure is employed to check the approximate constancy of the spectral determinant.
We finish by stating our final remarks in section \ref{Final}.

%\newpage

\section{$\mathcal{PT}$-symmetry in integrable field theories}\label{PTInt}

\qquad As we have argued, it is know for years that important models of classical field theory are associated to invariant equations under time reversal and parity operations.
In what follows we revisit some of them and highlight
%indicate/emphasize/illustrate/pinpoint/pointout/depict/ draw/callattentionto/explicitate/evidentiate* 
their behaviour under $\mathcal{PT}$.
More than that, we show that their most relevant solutions, namely their solitonic excitations, are also invariant.
Thus, the regime of stable particle-like waves which can scatter indefinitely, preserving the form of their profiles and conserving energy, coincide with the so called unbroken phase of $\mathcal{PT}$-symmetry.  

As introduced in \cite{BenderBoettcher}, the $\mathcal{PT}$-symmetry of a Hamiltonian system,
\begin{eqnarray}
[H,\mathcal{PT}]=0,
\end{eqnarray}
is not enough to guarantee the reality of its spectrum; 
instead one should also impose some restriction on the eigenvectors.
If the eigenstates are also invariant, up to a phase, under this symmetry,
\begin{eqnarray}
\mathcal{PT} \psi = \pm \psi
\end{eqnarray}
one is said to be in the unbroken $\mathcal{PT}$-symmetry regime.
It turns out that this concept can be transported to ensure that the energy of a specific field configuration in classical mechanics is real \cite{FringKdV}. The analogous expression for the unbroken $\mathcal{PT}$ symmetry condition is then
\begin{eqnarray}
%\mathcal{PT} \mathcal{H} = \mathcal{H}, \qquad
\mathcal{PT} u = \pm u.
\end{eqnarray}

Our findings suggest that some systems that are not exactly integrable but which satisfy an invariance of this type can behave as quasi-integrable, in the sense that they possess families of asympotically preserved charges.
Before studying the behaviour of $\mathcal{PT}$-symmetry with respect to deviations from the integrable domain, we present in the next subsections completely integrable systems and highlight their space-time invariances. 

\subsection{Korteweg - de Vries}

\qquad The KdV equation has been well studied also in the realm of non-Hermitian Hamiltonians. 
In \cite{BenderKdV}, it was shown that a certain $\mathcal{PT}$-symmetric deformation admits the existence of certain conservation laws
and in \cite{FringKdV} it has emerged that a family of Hamiltonian flows can be associated to a particular deformation of the KdV equation.
Further generalisations were considered in \cite{BenderCooper} and in \cite{AFCompactons} compactons were investigated.

But one does not have to impose $\mathcal{PT}$-symmetric extensions of the KdV in order to appreciate the effects of these simultaneous space-time transformations.
In fact, taking the standard equation, 
\begin{eqnarray}
u_{t} + a u u_{xx} + b u_{xxx}  =0, 
\end{eqnarray}
arising form the Hamiltonian density
\begin{eqnarray}
\mathcal{H} = -\frac{a}{6}u^3 + \frac{b}{2}u_x^2,
\end{eqnarray}
with usually $a=-6$ and $b=1$, one can see that the Hamiltonian remains ivariant if we replace $x \rightarrow - x$, $t \rightarrow - t$ and $u \rightarrow u$ at once.  

\begin{figure}[h!]
      \begin{center}
      \includegraphics[width=0.45\textwidth]{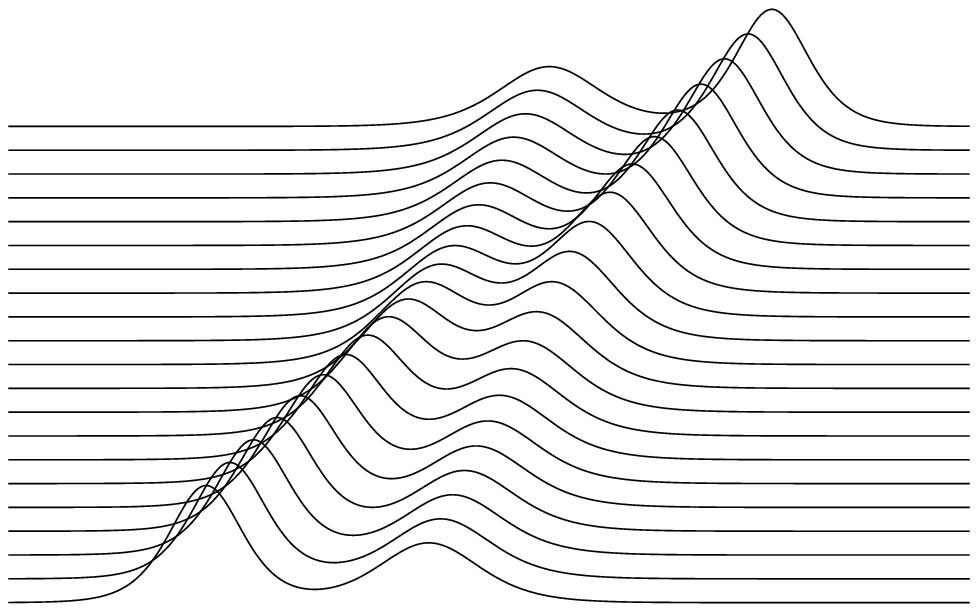}\label{figkdvr}
      \includegraphics[width=0.45\textwidth]{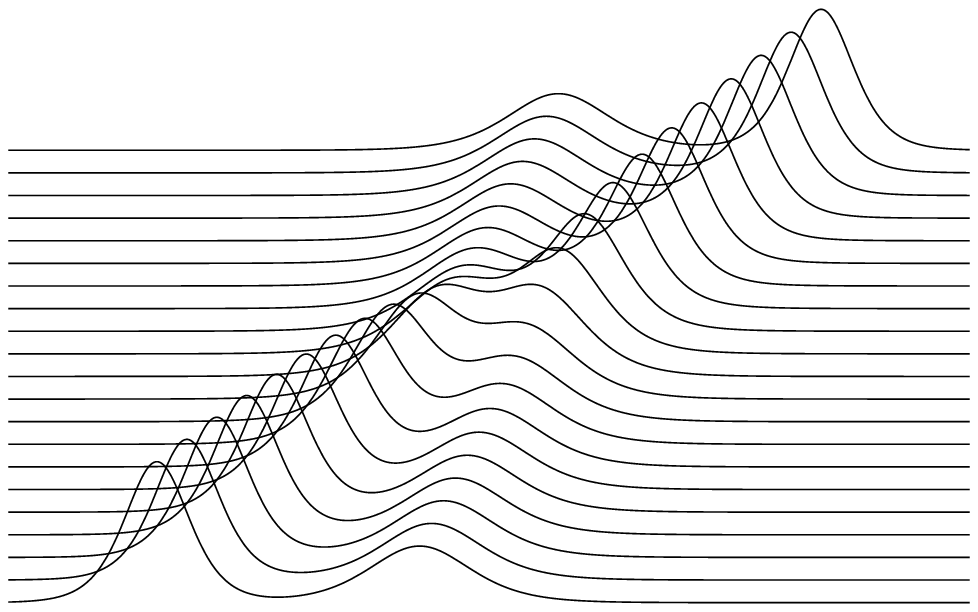}\label{figkdva}
      \caption{Colision of KdV solitons, repulsive (bounce-exchange) on the left and attractive (merge-split) on the right. The solution is $\mathcal{PT}$-symmetric with respect to the centre of mass reference frame.}\label{FigKdVs}
      \end{center}
      \end{figure}

This model admits multi-solitons excitations, and a two-soliton solution is known to be   
\begin{eqnarray}      
 u(x,t)=
	\frac{ 12\frac{c_1-c_2}{a b} 
	\left( c_2 \sinh ^2\left( \xi_1 \right)+
	       c_1 \cosh ^2\left( \xi_2 \right) \right) }
		{\left(
		\left(\sqrt{\frac{c_1}{b}}-\sqrt{\frac{c_2}{b}}\right) 
		\cosh\left( \xi_1 + \xi_2 \right)+
		\left(\sqrt{\frac{c_1}{b}}+\sqrt{\frac{c_2}{b}}\right) 
		\cosh \left( \xi_1 - \xi_2 \right)\right){}^2}, 
\end{eqnarray}
where we have abbreviated $\xi_i = \frac{1}{2} \sqrt{\frac{c_i}{b}} \left(x-c_i t\right)$.
The action of the symmetry operation into these objects gives $\mathcal{PT}[\xi_i] = -\xi_i$ making this solitonic solution $u(x,t)$, which is even in $\xi_i$, invariant, i.e. $\mathcal{PT}[u(x,t)] =u(-x,-t)=u(x,t)$, as required. In Figure \ref{FigKdVs} one can see the space-time symmetry of the two-solitons solutions.

These properties can be immediately identified if one plots the evolution surface for this KdV solitons collision, which we show bellow.
There is evidently an axis of symmetry with respect to which the simultaneous reflections on the space and time coordinates produce an identity operation, be it a repulsion or an attraction.

Like the KdV model, there are many other examples of integrable field equations  which not only present the parity-time reversal symmetry bu also admit $\mathcal{PT}$-symmetric stable solutions.
In what follows we discuss a few more illustrations.       

\subsection{Sinh-Gordon}

\begin{figure}[h!]
      \begin{center}
      \includegraphics[width=0.45\textwidth]{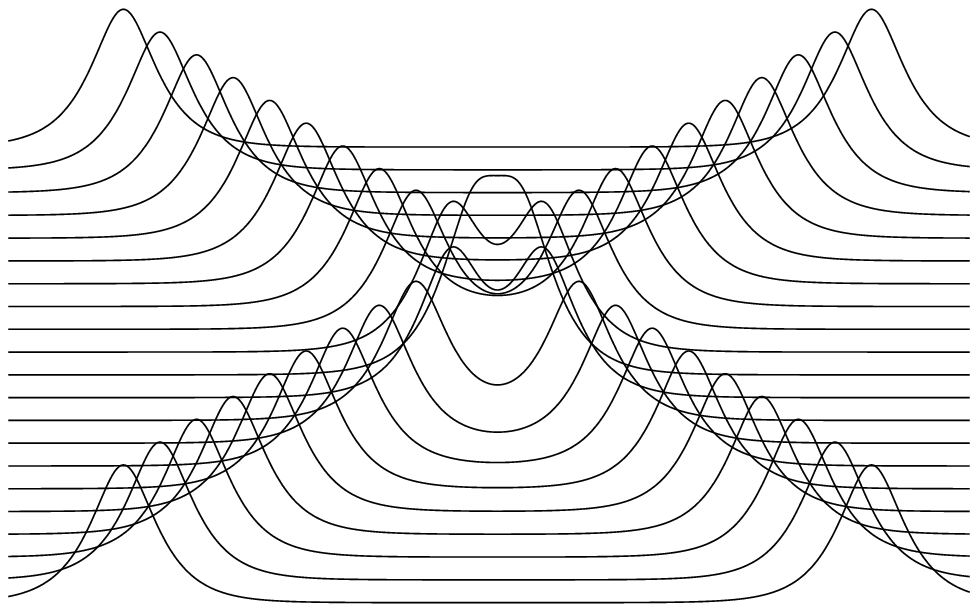}\label{figsolre}
      \includegraphics[width=0.45\textwidth]{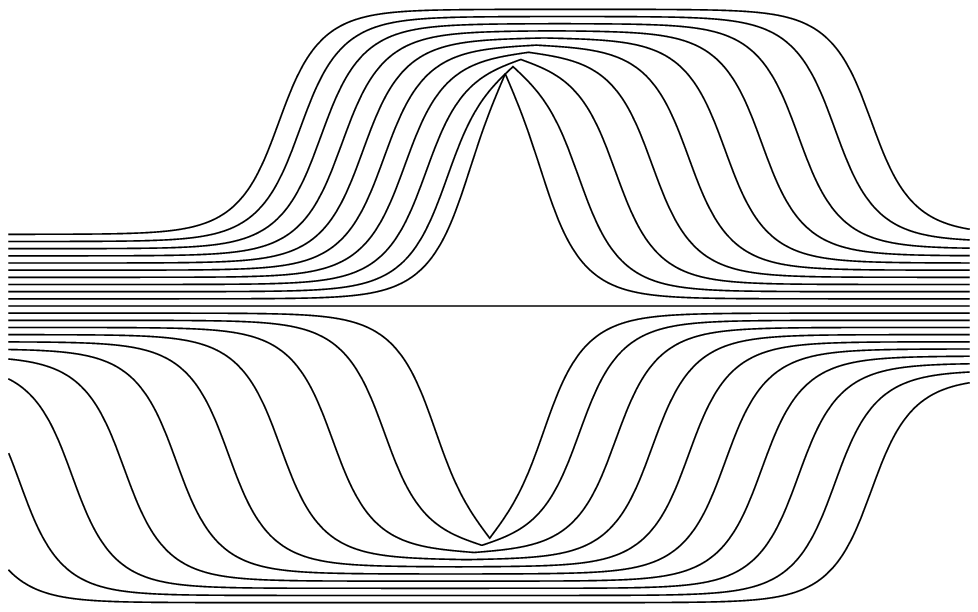}\label{figsolre}
      \caption{Sinh-Gordon complex solution (real part on the left, imaginary part on the right).}\label{FigShG}
      \end{center}
      \end{figure}

\qquad The Sinh-Gordon models is described by a Lorentz-invariant equation,
\begin{eqnarray}
\phi_{tt} - \phi_{xx} + \frac{\mu^2}{\beta}\sinh{\beta \phi}  =0, 
\end{eqnarray}
derived from
\begin{eqnarray}
\mathcal{H} = \frac{1}{2} \left( \phi_{t}^2 + \phi_{x}^2 \right) + \frac{\mu^2}{\beta^2}\left( \cosh{\beta \phi} -1 \right).
\end{eqnarray}
In the latter, the derivative terms are invariant under space-time reflections and all terms remain unchanged if solutions belong to the unbroken $\mathcal{PT}$-symmetry phase, $\mathcal{PT} \phi = \pm \phi$.
Remarkable solutions for this equation present a travelling wave form,  stable under collision, as depicted bellow. Some of these complex waves can be cast in the following form
%\begin{eqnarray}
%\phi (x,t,\omega,\mu) =
%\frac{4}{\beta }\tanh^{-1}\left(e^{\mu  (t \sinh (\theta )+x \cosh (\theta)+)+i \omega }\right),
%\end{eqnarray}
%\begin{eqnarray}
%\phi (x,t,\omega ,\mu ) &=& \frac{2}{\beta }\log \left(\frac{1-e^{\mu  \cosh (\theta ) (t \tanh (\theta )+x)+i \omega }}{1+e^{\mu  \cosh (\theta) (t \tanh (\theta )+x)+i \omega }}\right),
%\end{eqnarray}
%\begin{eqnarray}
%\varphi (x,t,\omega ,\mu ) &=& i \pi  \varphi (-x,-t,-\omega ,\mu ), \\
%\phi (x,t,\omega ,\mu ) &=& \phi (-x,-t,\omega ,-\mu ), 
%\end{eqnarray}
\begin{eqnarray}
\nonumber
\phi (x,t,\omega ,\mu ) &=& \frac{1}{\beta }\log \left(
   \frac{1-2 \cos (\omega ) e^{\mu  (t \sinh (\theta )+x \cosh (\theta ))}+e^{2 \mu  (t \sinh(\theta )+x \cosh (\theta ))}}
   {1+2 \cos (\omega ) e^{\mu  (t \sinh (\theta )+x \cosh (\theta ))}+e^{2 \mu  (t \sinh (\theta )+x \cosh (\theta))}}\right) \\
   &+& \frac{2 i}{\beta }\arctan\left(
   \frac{2 \sin (\omega ) e^{\mu  (t \sinh (\theta )+x \cosh (\theta ))}}
   {e^{2 \mu  (t \sinh(\theta )+x \cosh (\theta ))}-1}\right),
\end{eqnarray}
so the $\mathcal{PT}$-invariance of the equation of motion above, as $x \rightarrow -x$ and $t \rightarrow -t$, reads
\begin{eqnarray}
\phi (x,t,\omega ,\mu ) = \phi (-x,-t,-\omega ,\mu ) = \phi (-x,-t,\omega ,\mu )^{\ast},
%\phi \left(x,t,\frac{\pi }{2},\mu \right) &=& -\phi \left(-x,-t,\frac{\pi }{2},\mu \right).
\end{eqnarray}
in accordance with the complex conjugation that comes along with time-reversal in quantum, and complex, field theories (see Figure \ref{FigShG} for the real and imaginary parts of the excitation).
Here we have used $\theta = \tanh^{-1} \left( \frac{v}{c} \right)$ as the rapidity variable, with $c=1$.

\subsection{Sine-Gordon}      

\begin{figure}[h!]
      \begin{center}
      \includegraphics[width=0.5\textwidth]{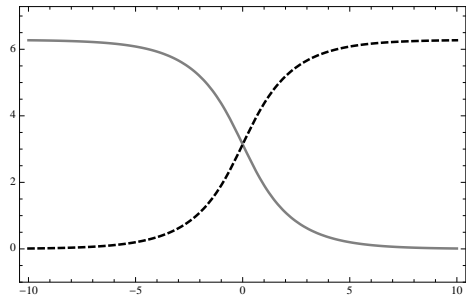}
      \caption{Sine-Gordon kink (dashed line, in black) and anti-kink (continuous line, in gray), interpolating neighbouring vacua.}\label{figkinks}
      \end{center}
      \end{figure}

\qquad By complexifying the coupling constant $\beta \rightarrow  i \gamma$ of the previous model, one obtains the Sine-Gordon Hamiltonian density,
\begin{eqnarray}
\mathcal{H} = \frac{1}{2} \left( \phi_{t}^2 + \phi_{x}^2 \right) - \frac{m^2}{\gamma^2}\left( \cos{\gamma \phi} -1 \right),
\end{eqnarray}
and its equation of motion,
\begin{eqnarray}
\phi_{tt} - \phi_{xx} + \frac{m^2}{\gamma}\sin{\gamma \phi}  =0, 
\end{eqnarray}
with localized solutions,
    \begin{equation}
     \phi(x,t) = \frac{4}{\gamma}\arctan\left( \epsilon e^{ m \cosh\theta \left( x + t \tanh\theta \right) } \right),
    \end{equation}
describing kinks ($\epsilon=+1$) and anti-kinks ($\epsilon=-1$), depicted in Figure \ref{figkinks}.
In order to appreciate its $\mathcal{PT}$ properties one should note that
 \begin{eqnarray}   
      4\left( \arctan\left( e^{\Gamma} \right) - \arctan\left(-e^{-\Gamma} \right) \right) = 0 \quad \textrm{mod} \quad 2\pi.
 \end{eqnarray}

An interesting feature is therefore that these two fundamental configurations, the kink and anti-kink, from which one can construct other excitations (breathers, bound states), are connected via $\mathcal{PT}$, up to a topological equivalence.
The system composed of a single (anti)-kink is not explicitly $\mathcal{PT}$-invariant but if one wants to investigate their stability one can create a collision between a kink and an anti-kink and this process is similar to the one shown in Figure \ref{FigShG} (on the right). Other non-symmetric collusions are also possible but in those cases topological arguments are to be used.

\subsection{Boussinesq equation and Calogero scattering}

\qquad Finally, we illustrate the good $\mathcal{PT}$ properties of integrable models by presenting one last example, that of the Boussinesq equation. In \cite{AFCalogero} we have explicitly constructed a class of solutions depending on dynamical movable poles.
Interestingly, the poles satisfy an interaction dictated by the Calogero Hamiltonian \cite{BoussCal}.
In that occasion we have concluded that $\mathcal{PT}$-symmetric Calogero particles would guarantee that the solutions to the Boussinesq equation have real energies.

Below we depict the evolution of the real and imaginary parts of the Boussinesq $\mathcal{PT}$-symmetric solution associated to the existence of three  Calogero-like poles.
      \begin{figure}[h!]
      \begin{center}
      \includegraphics[width=0.45\textwidth]{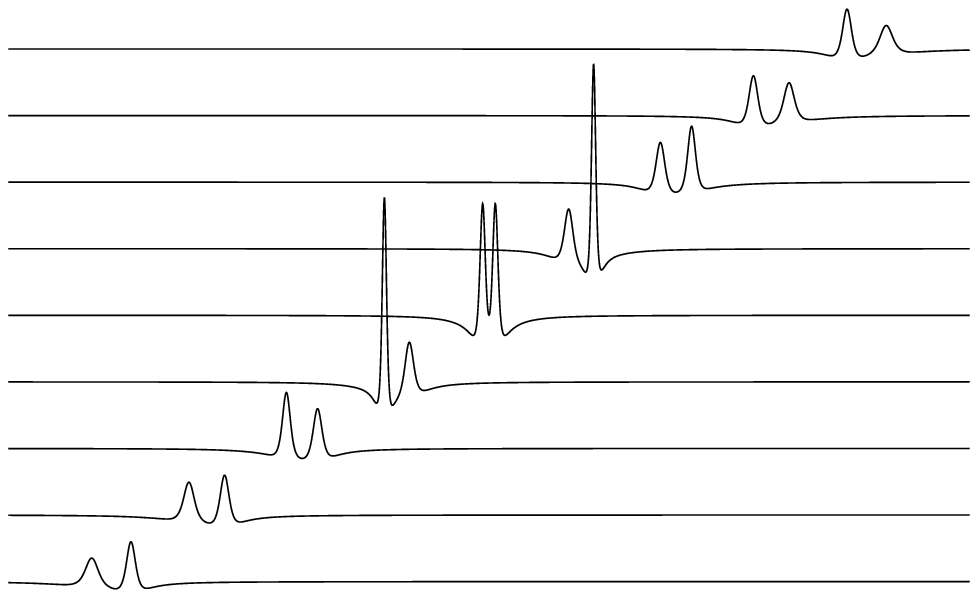}\label{figcalog}
      \includegraphics[width=0.45\textwidth]{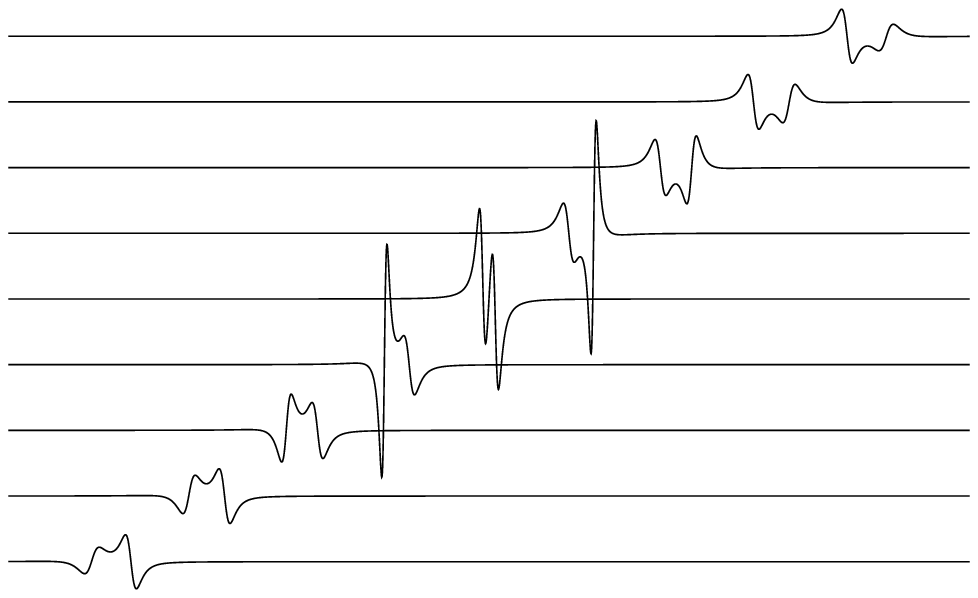}\label{figcaloge}
      \caption{Rational Boussinesq solitonic solution with Calogero poles: real (left) and imaginary (right) components, \cite{AFCalogero}.}\label{FigCalBouss}
      \end{center}
      \end{figure}
      
One can verify from Figure \ref{FigCalBouss} that the fields are indeed invariant under space and time reversal: the real part at two points $t$ and $-t$ correspond to a simple reflexion $x \rightarrow -x$, but the imaginary part also suffers a $-$ sign inversion, associated to the complex conjugation.
  
\begin{figure}[h!]
      \begin{center}
      \includegraphics[width=0.45\textwidth]{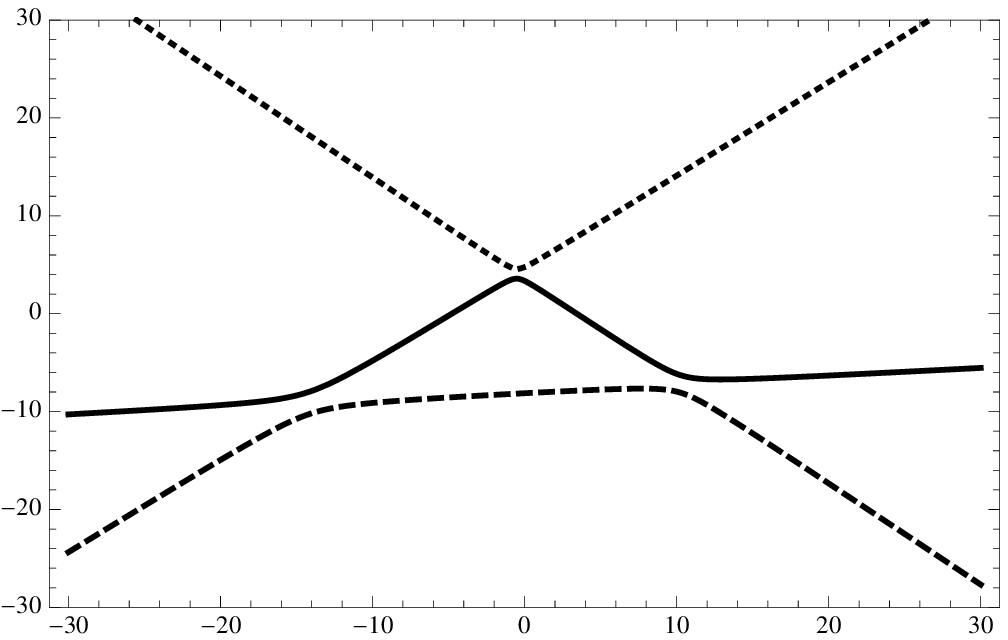}\label{figcalog}
      \includegraphics[width=0.45\textwidth]{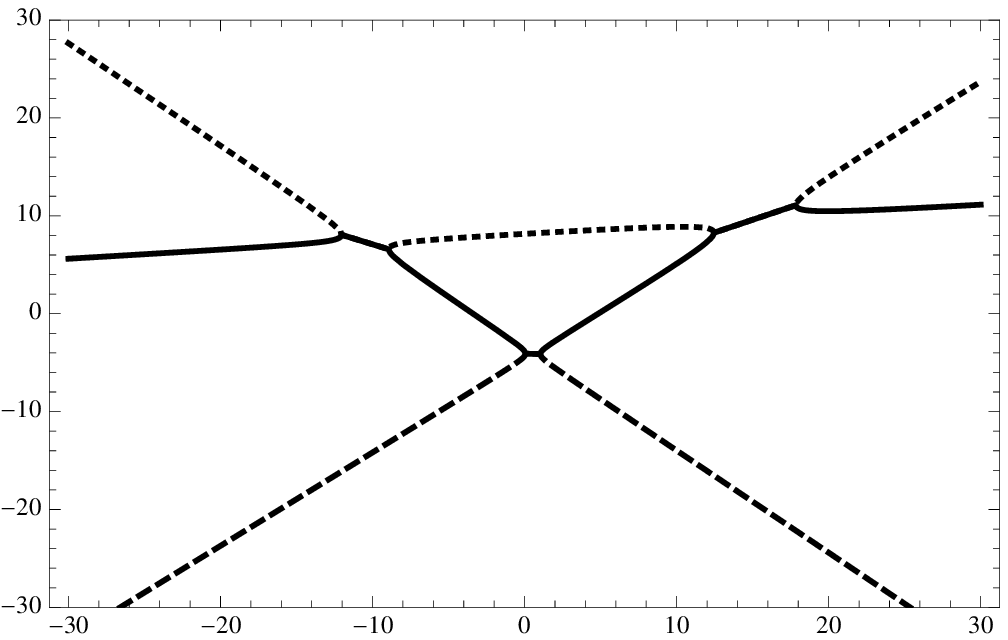}\label{figcaloge}
      \caption{Trajectories of a Calogero three-particle system in a classical scattering, 
      repulsive (bounce-exchange) on the left and attractive (merge-split) on the right, \cite{AFCalogero}.} \label{ColiCalo}
      \end{center}
      \end{figure}

The poles appearing in this solution interact as shown in Figure \ref{ColiCalo}.
If we decompose the scattering in two-bodies collisions, the trajectories of the particles are symmetric with respect to space-time inversions relative to the centre of mass reference frame.

%\subsection{Scattering properties}            
%\qquad

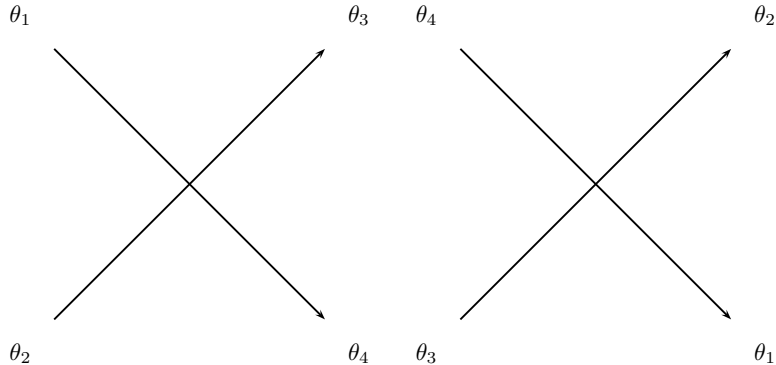
\begin{figure}[h]
\psscalebox{0.9}{
\begin{pspicture}(0,0)(0,0)
\psline{->}(2,-5)(6,-1)
\psline{->}(2,-1)(6,-5)
\psline{->}(8,-5)(12,-1)
\psline{->}(8,-1)(12,-5)
\rput(1.5,-0.5){$\theta_1$}
\rput(1.5,-5.5){$\theta_2$}
\rput(6.5,-0.5){$\theta_3$}
\rput(6.5,-5.5){$\theta_4$}
\rput(12.5,-0.5){$\theta_2$}
\rput(12.5,-5.5){$\theta_1$}
\rput(7.5,-0.5){$\theta_4$}
\rput(7.5,-5.5){$\theta_3$}
\end{pspicture}
}
\vspace{5cm}
\caption{Two-particle scattering invariant under $\mathcal{PT}$ transformation, up to a minor relabeling.}\label{Scat}
\end{figure}

Although constructed from a classical starting point, these ideas can be  extended to quantum scattering processes.
The classical collisions above resemble the decomposition property of the scattering matrix.
From Figure \ref{Scat} we see the effect of space-time reversal and it becomes evident that indeed $\mathcal{PT}{}^2=1$.

In the case of quasi-integrable systems, despite the asymptotic stability of the excitations one observes the presence of radiation near the interaction region.
It has been noticed \cite{Zwanziger, Miramontes,OlallaFringUnstable} that unstable particles  can be produced and annihilated without being observed asymptotically in a scattering process. These particles are associated to complex poles in the scattering amplitude and could be related to the present observations.

%\subsubsection{Quantum integrable $\mathcal{PT}$-symmetric scattering} 
%\newpage    
%\begin{figure}[h]
%\begin{pspicture}(0,0)(0,0)
%\psline{->}(2,-5)(6,-1)
%\psline{->}(2,-1)(6,-5)
%\psline{->}(8,-5)(12,-1)
%\psline{->}(8,-1)(12,-5)
%\rput(1.5,-0.5){$\theta_1$}
%\rput(1.5,-5.5){$\theta_2$}
%\rput(6.5,-0.5){$\theta_3$}
%\rput(6.5,-5.5){$\theta_4$}
%\rput(12.5,-0.5){$\theta_1$}
%\rput(12.5,-5.5){$\theta_2$}
%\rput(7.5,-0.5){$\theta_3$}
%\rput(7.5,-5.5){$\theta_4$}
%\end{pspicture}
%\vspace{6cm}
%\caption{Two-particle scattering invariant under $\mathcal{PT}$ transformation (relabeling?).}
%\end{figure}

\section{Path-independence and conservation laws}\label{Cons} 

\qquad Once appreciated the role of $\mathcal{PT}$-symmetry in integrable systems, we now turn our attention to an attempt to understand the origin of  important associated properties, more notably the stability of solutions. In order to do so, we must start from a formulation of integrable systems in which the space and time coordinates arise naturally.
In fact, an integrable theory can be formulated in terms of Wilson loops in space-time and the non abelian Stokes theorem \cite{Alvarez},
\begin{eqnarray}
W(\Gamma) \equiv 
\hat{\mathcal{P}}\exp 
\Bigg[ - \oint_\Gamma A_\mu dx^\mu \Bigg] =
\hat{\mathcal{P}}\exp 
\Bigg[ - \int_{\Sigma} d x^\nu d x^\mu  W^{-1} F_{\mu\nu} W \Bigg] \equiv
V(\Sigma).\label{VW}
\end{eqnarray}

\begin{figure}[h]
\psscalebox{0.9}{
\begin{pspicture}(0,0)(1,0)
\psline{->}(7.5,-9)(7.5,-1)\rput(7.5,-0.5){$t$}
\psline{->}(2,-5)(13,-5)\rput(13.5,-5){$x$}
\psline[linestyle=dashed]{->}(3,-8)(12,-8)\rput(8,-8.5){$-\tau$}\rput(9,-7.5){$W(\Gamma_{-\tau})$}
\psline[linestyle=dotted]{->}(12,-8)(12,-2)\rput(12.5,-4.5){$+L$}\rput(11,-4.5){$W(\Gamma_{+L})$}
\psline[linestyle=dashed]{->}(12,-2)(3,-2)\rput(8,-1.5){$+\tau$}\rput(6,-2.5){$W(\Gamma_{+\tau})$}
\psline[linestyle=dotted]{->}(3,-2)(3,-8)\rput(2.5,-4.5){$-L$}\rput(4,-5.5){$W(\Gamma_{-L})$}
\psline[linestyle=dashed,dash=3pt 2pt]{-}(2.1,-8.55)(12.9,-1.45)
\rput(12,-1){$\mathcal{PT}$-symmetry axis}
\end{pspicture}}
\vspace{9cm}
\caption{A closed path will in general produce a phase shift in (\ref{psiaux}) due to the presence of the potentials $A_\mu$ in that region of space. However, if these gauge potentials present certain behaviour under parity reflection and time reversal then the integral in (\ref{phaseintpath}) can be path-independent.} \label{FigXT}
\end{figure}
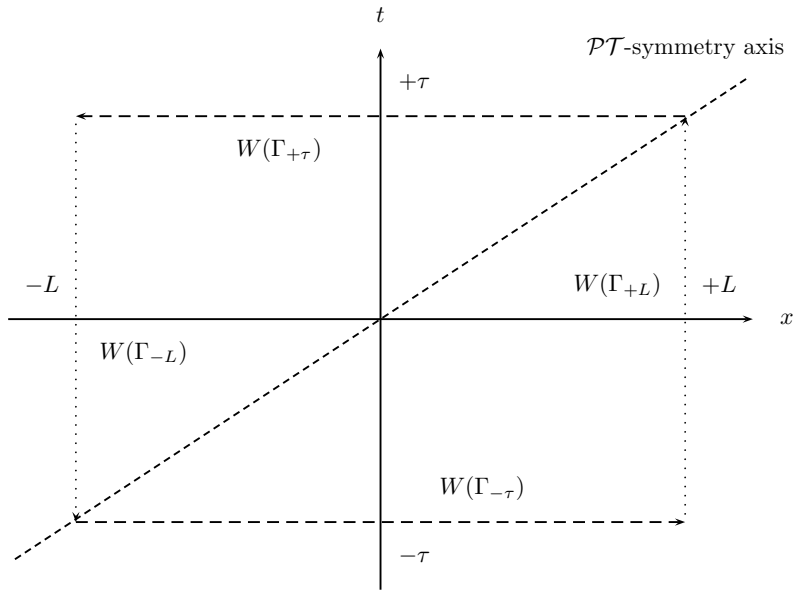

One can have closed paths starting (and finishing) at different points.
For instance, using the four vertices - $(-L,-\tau), (+L,-\tau),(+L,+\tau), (-L,+\tau)$ - of Figure \ref{FigXT}, we can write, for a closed contour around the point $(-L,-\tau)$,
\begin{eqnarray}
W(\Gamma_{C_-}) = W(\Gamma_{-L}) W(\Gamma_{+\tau}) W(\Gamma_{+L}) W(\Gamma_{-\tau}).\label{WWWW}
\end{eqnarray}

We note that the object $W$ in (\ref{VW}) will depend on
\begin{eqnarray}
\int dt A_t(L,t) + \int dt A_t(-L,t)
=
\int dt A_t(L,t) - \int dt A_t(-L,-t)
\end{eqnarray}
and
\begin{eqnarray}
\int dx A_x(x,-\tau) + \int dx A_x(x,\tau)
=
-\int dx A_x(-x,-\tau) + \int dx A_x(x,\tau)
\end{eqnarray}
which will vanish if these gauge potentials are invariant under parity and time reversal operations,
\begin{eqnarray}
A_t(L,t) = A_t(-L,-t), \label{Atsym} \\
A_x(x,\tau) = A_t(-x,-\tau). \label{Axsym}
\end{eqnarray}

Indeed, 
%if
%\begin{eqnarray}
%A_t(+L,t) = A_t(-L,t), \\ 
%A_x(x,+\tau) = A_t(x,-\tau), 
%\end{eqnarray}
then we have, respectively,
\begin{eqnarray}
U(t) = W(\Gamma_{+L}) = W(\Gamma_{-L})^{-1}, \\ 
S(x) = W(\Gamma_{+\tau}) = W(\Gamma_{-\tau})^{-1},
\end{eqnarray}
and consequently
\begin{eqnarray}
W(\Gamma_{C_-}) = U(t)^{-1} S(x) U(t) S(x)^{-1}, 
\end{eqnarray}
can be associted to a conserved object, since
\begin{eqnarray}
\det W(\Gamma_{C_-}) =  1.
\end{eqnarray}

If we have an auxiliary vector $\psi$ of the following form
\begin{eqnarray}\label{psiaux}
\psi(x,t) = W(\Gamma) \psi_0,
\end{eqnarray}
with
\begin{eqnarray}\label{phaseintpath}
W(\Gamma) = \hat{\mathcal{P}}\exp \left( \int_{\Sigma(\Gamma)} dx \, dt\, W(x,t)^{-1} F_{xt}(x,t) W(x,t) \right),
\end{eqnarray}
where the nonintegrable phase of this path-ordered exponential is dictated by the object $F_{xt}(x,t)$, then its value will heavily rely on the chosen curve $\Gamma$.
If $F_{xt}(x,t)$ vanishes, the system becomes integrable as the phase trivialises and the auxiliary vector satisfies
\begin{eqnarray}
\left[ ~ \partial_t + A_t(x,t) ~\right] \psi(x,t) &=& 0, \label{SturmT}\\
\left[ \, \partial_x + A_x(x,t) ~\right] \psi(x,t) &=& 0, \label{SturmX}
\end{eqnarray}
and the object $F_{xt}(x,t)$ can be identified with
\begin{eqnarray}
F_{xt} \equiv [\partial_t + A_t, \partial_x + A_x] = 0.
\end{eqnarray}

If one is concerned with closed paths $\Gamma_C$ there is another way of obtaining a trivial integrated phase, $W(\Gamma_C) =1$,
and that is if the integrand is symmetric with respect to space and time along a certain integration contour, mimicking the effects of a zero curvature condition 
$F_{xt}=0$.
That means that, albeit not being necessarily integrable, a model might present excitation configurations that resemble in some senses those of an integrable theory.

So far we have presented various examples of integrable systems, real and complex, and we have highlighted their $\mathcal{PT}$-symmetric properties.
In what follows we somehow reverse the logic and present some examples of nonintegrable models and argue that if $\mathcal{PT}$-symmetry is present then we can
appreciate the survival of a few features which are reminiscent of integrable systems.

To achieve this we have chosen a deformation of the sine-Gordon model which was introduced in \cite{Bazeia} and further explored in \cite{FerrWojZak}, where stability and symmetry properties have been spotted.

\section{The $\mathcal{PT}$-integrabitliy of a deformed sine-Gordon}\label{BSG}

\begin{figure}[h!]
\begin{center}
\includegraphics[width=0.45\textwidth]{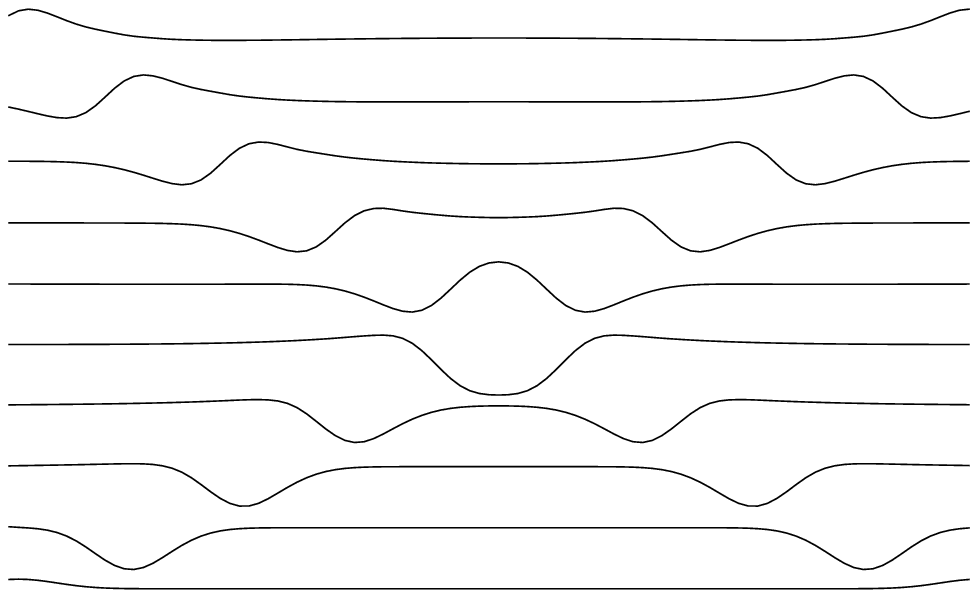}
\includegraphics[width=0.45\textwidth]{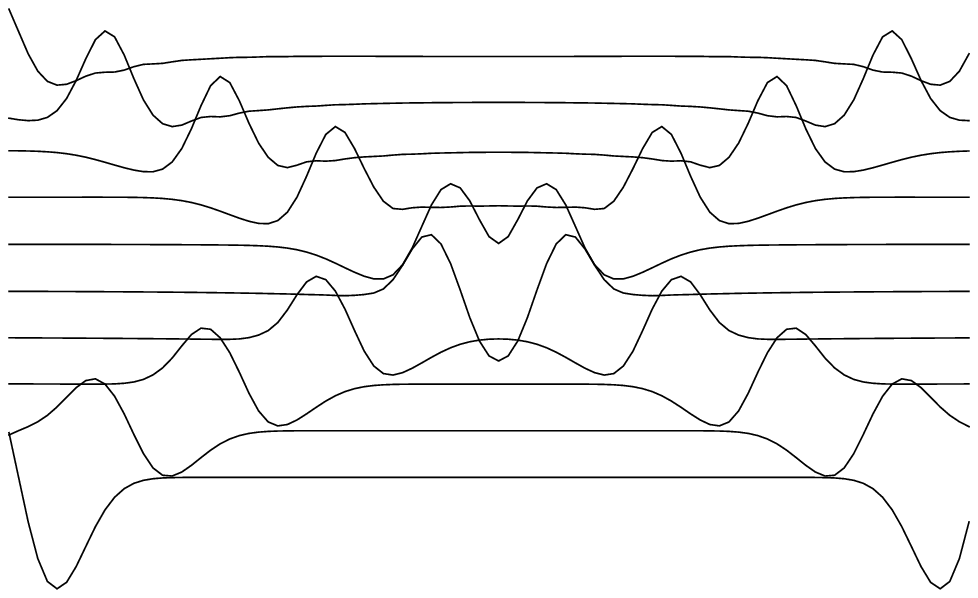}\\
\includegraphics[width=0.45\textwidth]{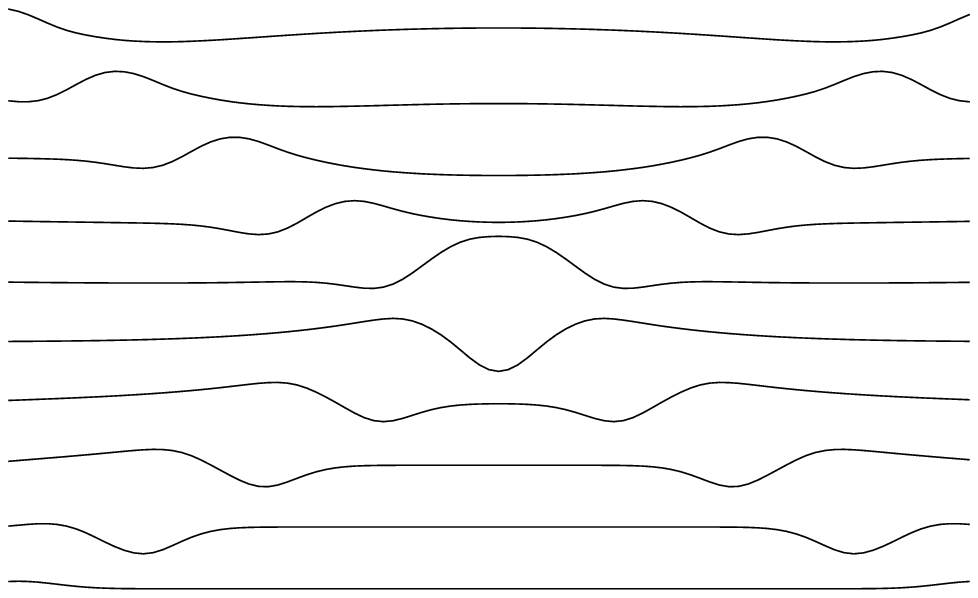}
\includegraphics[width=0.45\textwidth]{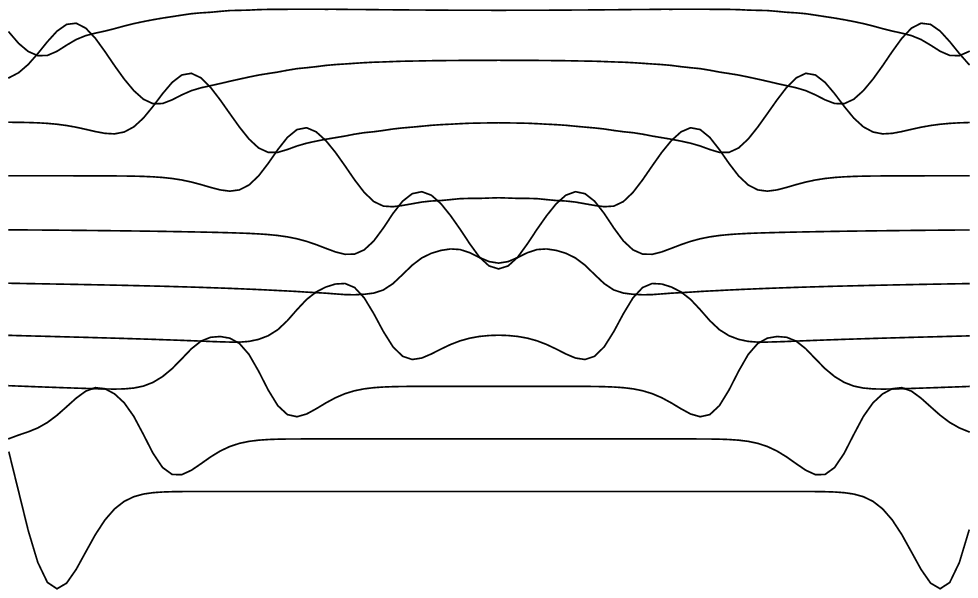}\\
\includegraphics[width=0.45\textwidth]{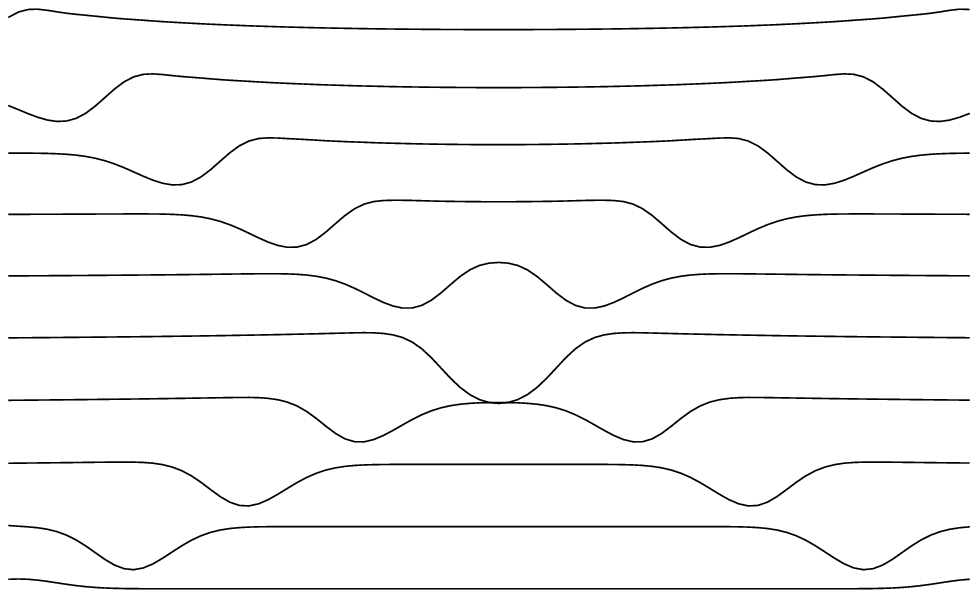}
\includegraphics[width=0.45\textwidth]{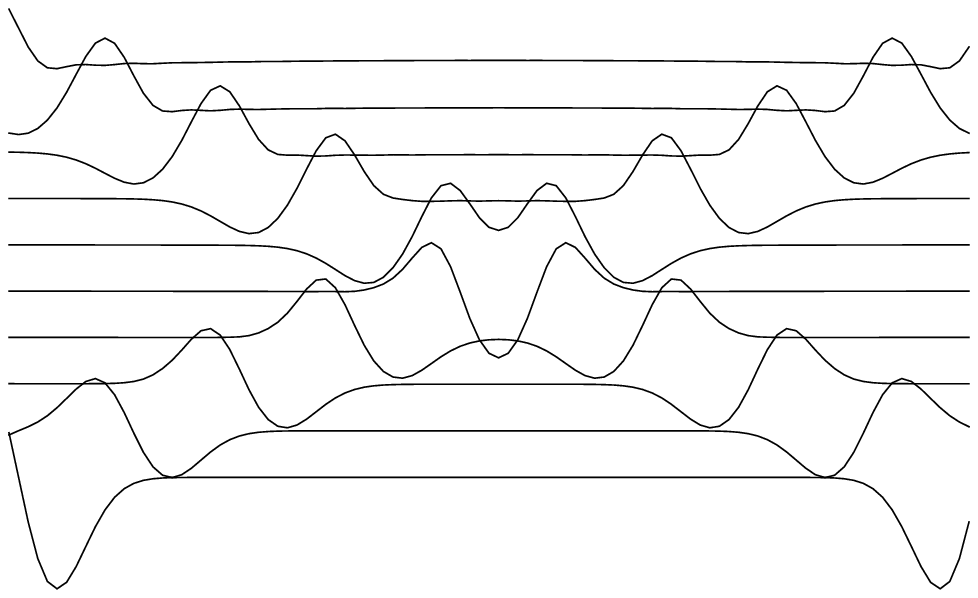}\\
\includegraphics[width=0.45\textwidth]{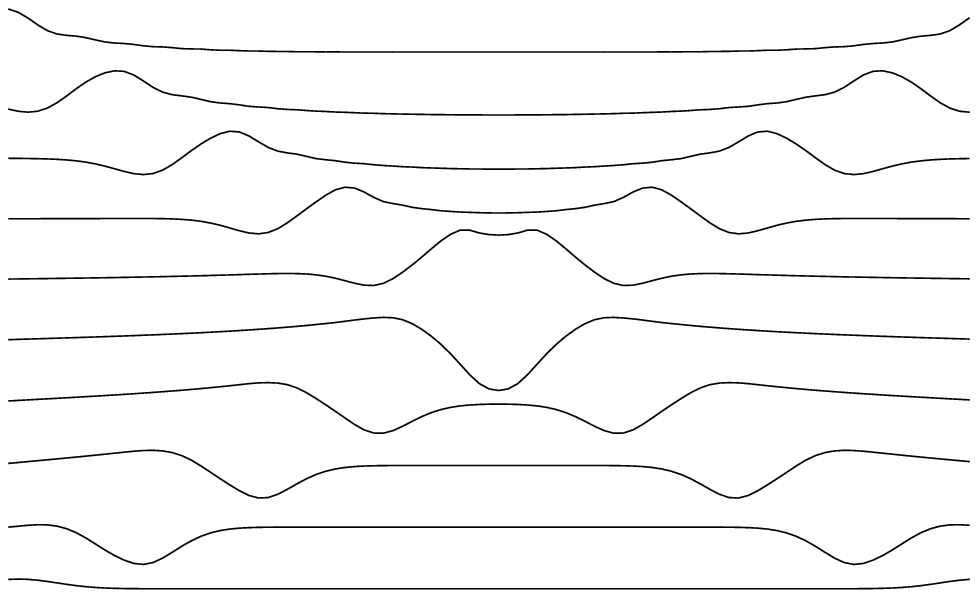}
\includegraphics[width=0.45\textwidth]{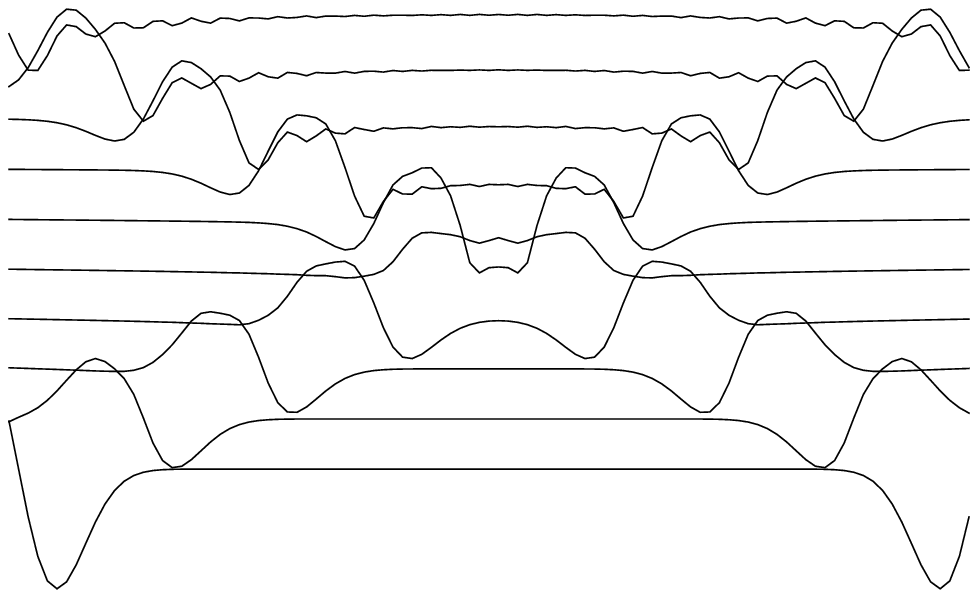}
\caption{Solitary-like waves collision for n=2,3,4,5 depicted in terms of the fields $\Phi$ (left) and momenta $\Pi$ (right).
As the deformation parameter $\varepsilon=n-2$ increases, the less stable the scattering tends to be and some radiation starts to be observed after the collision, breaking the time reversal symmetry. There is also a slightly better behaviour of positive $n$ compared to odd values. }\label{SolWaves}
\end{center}
\end{figure}

\qquad Taking specifically the example of the sine-Gordon model,
we can write the gauge fields associated to the laboratory coordinates as in \cite{FerrWojZak}.
For our purposes here, in investigating the Inverse Scattering problem for quasi-integrable models, it will be more convenient to express the problem in terms of space and time coordinates, rather than light-cone coordinates. Starting from a standard form of a more symmetric Lax pair associated with the sine-Gordon model, we express it in terms of a generic potential $V(\Phi)$. The result are the following gauge fields,
\begin{eqnarray}
A_t(x,t)= \frac{1}{2i} B  \Phi_x (x,t)\sigma_z + i\alpha_+, \label{AtPot}\\
A_x(x,t)= \frac{1}{2i} B  \Phi_t (x,t)\sigma_z + i\alpha_-, \label{AxPot}
\end{eqnarray}
with
\begin{eqnarray}
\alpha_\pm = \frac{1}{2 \sqrt{2}}\left( \pm \frac{1}{2}
\left(\Lambda \pm \frac{1}{\Lambda}\right)\frac{V'(\Phi(x,t))}{\sqrt{V(\Phi(x,t))}}\sigma_x\mp B\left(\Lambda \mp \frac{1}{\Lambda}\right)\sqrt{V(\Phi(x,t))}\sigma_y \right), 
\end{eqnarray}
making the $A_\mu$ potentials invariant under the operations of space-time reversal, (\ref{Atsym}) and (\ref{Axsym}), if so are the fields.

As a matter of fact, these properties also render the equations (\ref{SturmT}) and (\ref{SturmX}) $\mathcal{PT}$-symmetric if one considers the parity transformation to act on the $2 \times 2$ Pauli matrices as
$\mathcal{P} = 
\left(
\begin{array}{cc}
0 & 1 \\
1 & 0 
\end{array}
\right)$,
\begin{eqnarray}
\begin{array}{ccccccccccc}
\mathcal{P} \sigma_x \mathcal{P} & = & \sigma_x, 
& \quad & 
\mathcal{T} \sigma_x \mathcal{T} & = & \sigma_x,
& \quad &
\mathcal{PT} \sigma_x \mathcal{PT} & = & \sigma_x,\\
\mathcal{P} \sigma_y \mathcal{P} & = & -\sigma_y, 
& \quad & 
\mathcal{T} \sigma_y \mathcal{T} & = & -\sigma_y,
& \quad &
\mathcal{PT} \sigma_y \mathcal{PT} & = & \sigma_y,\\
\mathcal{P} \sigma_z \mathcal{P} & = & -\sigma_z, 
& \quad & 
\mathcal{T} \sigma_z \mathcal{T} & = & \sigma_z,
& \quad &
\mathcal{PT} \sigma_z \mathcal{PT} & = & -\sigma_z.
\end{array}
\end{eqnarray}
The unbroken $\mathcal{PT}$-symmetry of the classical field is thus not directly related to the unbroken phase of the vector $\psi$, but with the $\mathcal{PT}$-symmetry of the equation itself. 

%Similarly, the same is true if the gauge potentials are $\mathcal{PT}$ symmetric.

%, $\mathcal{PT} A_\mu(x,t) = A_\mu(-x,-t)^\ast = A_\mu(x,t) $.
The form of the $A_\mu$ potentials in (\ref{AtPot})-(\ref{AxPot}) is convenient as it allows one to treat nonintegrable extensions, such as \cite{Bazeia}
\begin{eqnarray}\label{poteBazeiaN}
V(\phi) = \frac{32}{n^2} \frac{(4 M)^2}{(2 B)^2} \tan^2{ \left( \frac{B \phi} {2} \right)} \left(1 - \sin^n{\left( \frac{B \phi}{2}\right)} \right)^2,
\end{eqnarray}
depending on a deformation parameter $\varepsilon=n-2$, which vanishes for the the sine-Gordon case, with $n=2$.
%which satisfy $L_i^\ast = \sigma_y \, L_i \, \sigma_y$ and
Such gauge connections correspond to an anomalous curvature,
\begin{eqnarray}
F_{xt} &=& \left( \Phi_{tt}(x,t)-\Phi_{xx}(x,t) + V'(\Phi (x,t)) \right) \sigma_z + \Delta, \\[10pt]
\nonumber
\Delta &=& \frac{1}{2 \sqrt{2} B}\left(\frac{1}{2}\left(\Lambda -\frac{1}{\Lambda }\right)\Phi_t (x,t)+
\frac{1}{2}\left(\Lambda +\frac{1}{\Lambda }\right)\Phi_x (x,t)\right) \times \\
&\times&
\left(4 B^2 \sqrt{V(\Phi (x,t))}+\frac{2 V''(\Phi (x,t))}{\sqrt{V(\Phi (x,t))}}-\frac{V'(\Phi (x,t))^2}{V(\Phi (x,t))^{3/2}}\right)\sigma_x.
\end{eqnarray}

It is interesting to note that, again, the zero curvature condition depends not only on the model - as it is the case for integrable systems - but also on the solution.
However, in order to have a vanishing anomaly one needs constant solutions, $\Phi_t (x,t)=\Phi_x (x,t)=0$, which necessarily occurs asymptotically for models with degenerate vacua.
Unless, of course, if $V=V_{SG}$, when the Lax pairs are those commonly found in the literature, e.g. \cite{BabelonBernardTalon}, and the second factor of $\Delta$ is identically satisfied.

From Figure \ref{SolWaves} indicate that as the deformation parameter increases there is a tendency of forming radiation waves, as observed in \cite{FerrWojZak}.
These excitations cannot be identified with unstable particles \cite{Zwanziger,Miramontes,OlallaFringUnstable}, as the latter are present in integrable systems and the former constitute a sign of the breaking of integrability.

\begin{figure}[h]
\begin{center}
\includegraphics[width=0.47\textwidth]{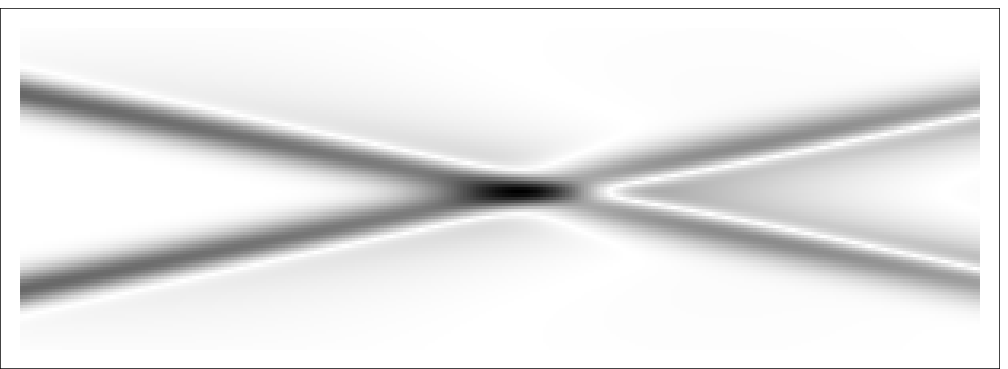}
\includegraphics[width=0.47\textwidth]{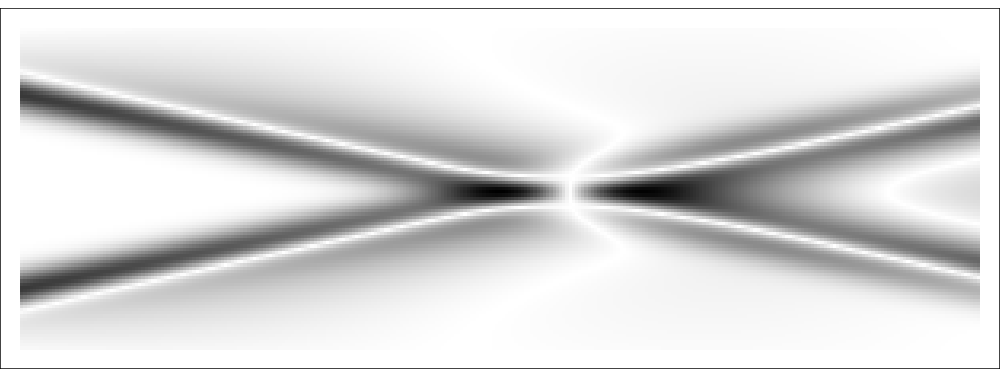}
\\
\includegraphics[width=0.47\textwidth]{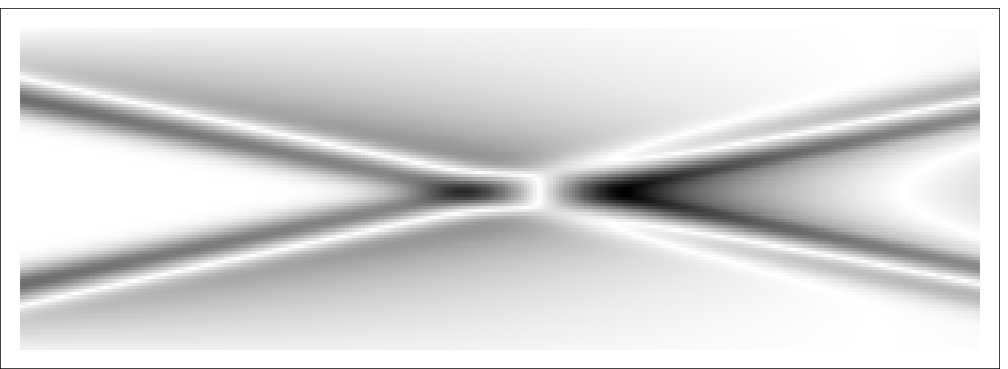}
\includegraphics[width=0.47\textwidth]{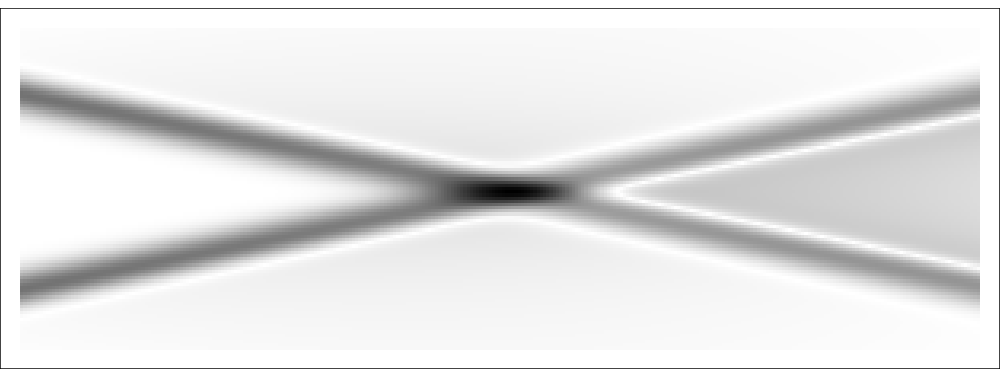}
\caption{Solitary waves scattering for the potential in equation (\ref{poteBazeiaN}) with $n=2,3,4,5$ clockwise from top left. In the vertical axes we have plot the space coordinates (x) and in the horizontal line the time coordinates (t) flow; the amplitude of the wave at each point is representated in greyscale, for which the larger amplitudes are plotted in darker shades.}\label{SolWaveScatPlot}
\end{center}
\end{figure}

\section{Symmetries of spectral properties}\label{Spec}

\qquad An important property of the spectral coefficients $a(\Lambda)$ and $b(\Lambda)$ is that the action-angle variables are constructed in terms of them.
This implies one can expand the action variable in powers of the spectral parameter $\Lambda$, so that if action variable is constant then one can generate infinitely-many conserved charges. Relaxing this latter condition a bit, if the action variable changes slowly with time keeping its final state equal to its initial state, then we have an infinite number of quantities whose values are asymptotically the same. 

 \begin{figure}[h!]
 \begin{center}
 \includegraphics[width=0.45\textwidth]{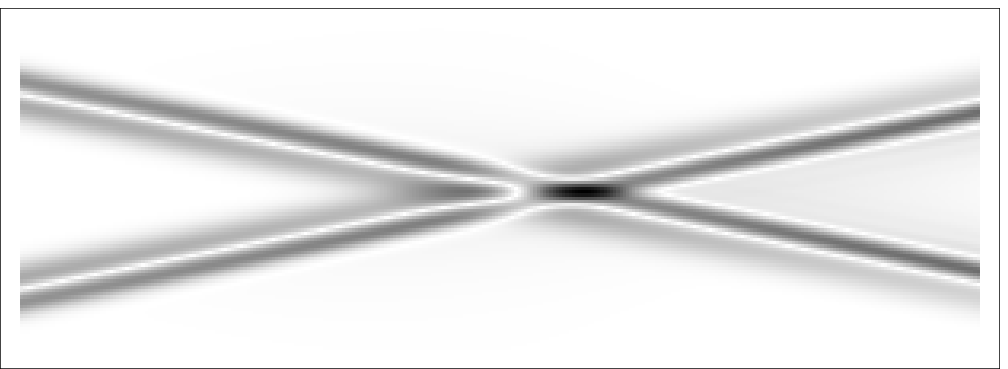}
 \includegraphics[width=0.45\textwidth]{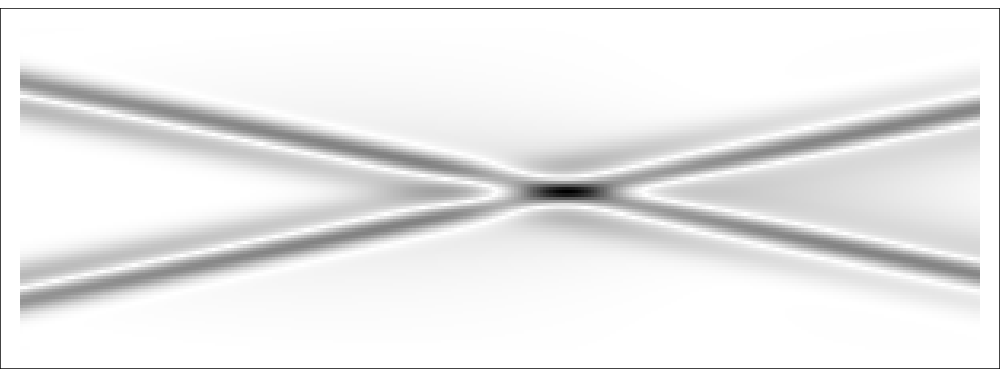}\\
 \includegraphics[width=0.45\textwidth]{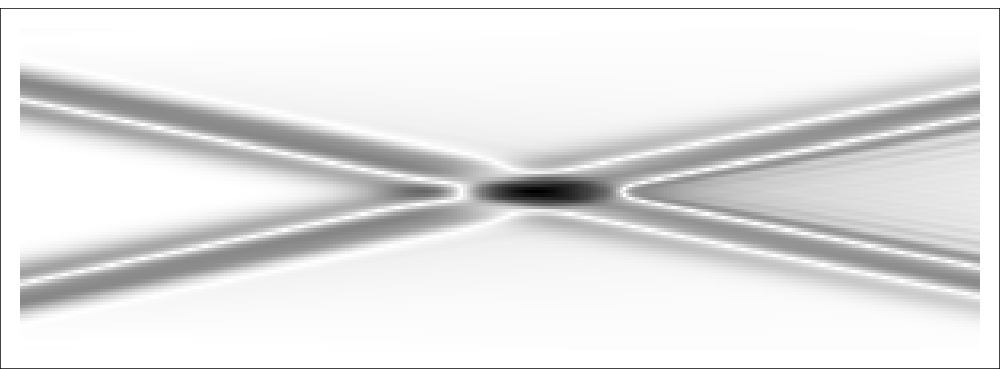}
 \includegraphics[width=0.45\textwidth]{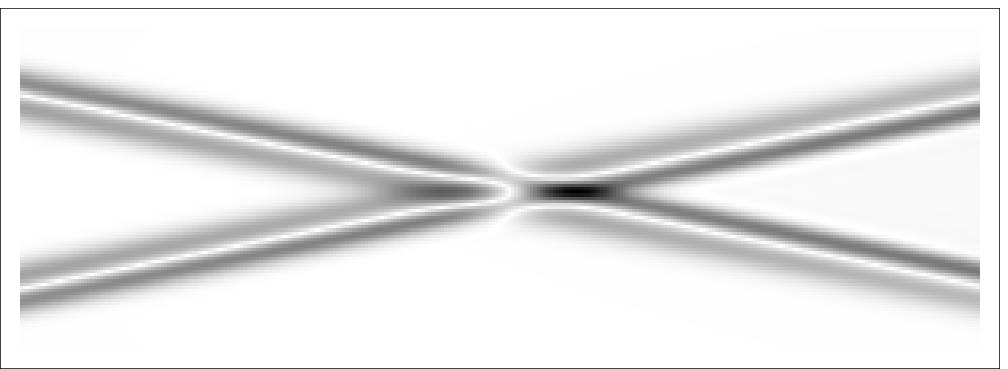}
 \caption{Diagonal components of $A_x(x,t)$ for interaction potential in equation (\ref{poteBazeiaN}) with $n=2,3,4,5$ clockwise from top left.
 Similar figures are produced when computing the off-diagonal components of $A_x(x,t)$, as well as diagonal and off-diagonal components of $A_t(x,t)$}\label{AxD}
 \end{center}
 %\end{figure}
 %\begin{figure}[h]
% \begin{center}
% \includegraphics[width=0.4\textwidth]{QSGLX12N2.eps}
% \includegraphics[width=0.4\textwidth]{QSGLX12N3.eps}\\
% \includegraphics[width=0.4\textwidth]{QSGLX12N5.eps}
% \includegraphics[width=0.4\textwidth]{QSGLX12N4.eps}
% \caption{Off-diagonal components of $A_x(x,t)$ for interaction potential in equation (\ref{poteBazeiaN}) with $n=2,3,4,5$ clockwise from top left.}\label{AxOD}
% \end{center}
% \end{figure}
% \begin{figure}[h!]
% \begin{center}
% \includegraphics[width=0.4\textwidth]{QSGLT11N2.eps}
% \includegraphics[width=0.4\textwidth]{QSGLT11N3.eps}\\
% \includegraphics[width=0.4\textwidth]{QSGLT11N5.eps}
% \includegraphics[width=0.4\textwidth]{QSGLT11N4.eps}
% \caption{Diagonal components of $A_t(x,t)$ for interaction potential in equation (\ref{poteBazeiaN}) with $n=2,3,4,5$ clockwise from top left.}\label{AtD}
% \end{center}
 %\end{figure}
 %\begin{figure}[h]
% \begin{center}
% \includegraphics[width=0.4\textwidth]{QSGLT12N2.eps}
% \includegraphics[width=0.4\textwidth]{QSGLT12N3.eps}\\
% \includegraphics[width=0.4\textwidth]{QSGLT12N5.eps}
% \includegraphics[width=0.4\textwidth]{QSGLT12N4.eps}
% \caption{Off-diagonal components of $A_t(x,t)$ for interaction potential in equation (\ref{poteBazeiaN}) with $n=2,3,4,5$ clockwise from top left.}\label{AtOD}
% \end{center}
 \end{figure}

Thus, if one is concerned only with the in- and out-states in a far past and a far future, respectively, there is the possibility of constructing a great deal of objects which are conserved in a broader sense. In order to so, it is essential to investigate the behaviour of the spectral data, $a(\Lambda)$ and $b(\Lambda)$.
We can expand at $x\rightarrow \infty$
\begin{eqnarray}\label{spectralexpansionvector}
\psi_+(x,\Lambda) = a(\Lambda) \overline{\psi_-}(x,\Lambda)+b(\Lambda) \psi_-(x,\Lambda),
\end{eqnarray}
and apply to it the time evolution.

As a consequence, one can show that, if the phases tend to multiple values of %$\pi \, Q$ ($Q$ being the associated 
the asymptotic vacua topological charge, then the left hand side of the first equation identically vanishes and
we automatically have that the $a(\Lambda)$ keeps its original value, $a_0(\Lambda)$, if one is interested in asymptotic states only.
On the other hand, as can be seen in more detail in \cite{AssisFerreiraQSG}, the other spectral coefficient  evolves according to the potential $V(\Phi)$, evaluated at the initial vacua configurations $\Phi_-$, 
%\begin{eqnarray}
%\partial_t \; a(\Lambda ,t) &=& 0, \\
%\partial_t \; b(\Lambda ,t) &=& \frac{i}{4 \sqrt{2} M}
%\left(
%\omega (\Lambda )\frac{V'(\Phi_-)}{\sqrt{V(\Phi_-)}}
%-2i\,B \, k(\Lambda)\sqrt{V(\Phi_-)}
%\right)   b(\Lambda ,t),
%\end{eqnarray}
%or, for $t \rightarrow \infty$ \footnote{This conservation is not valid for any instant $t$ because we know that in general the covariant derivatives do not commute and intermediately the curvature does not vanish.},
\begin{eqnarray}
a(\Lambda ,t) = a_0(\Lambda) , \qquad
b(\Lambda ,t) = b_0(\Lambda) \, e^{i \, \gamma \, t} \, e^{\Gamma \, t}, 
\end{eqnarray}
with an oscillating phase, 
\begin{eqnarray}
\gamma &=& 
\frac{\omega (\Lambda)}{4 \sqrt{2} M}  \frac{V'\left(\Phi _-\right)}{\sqrt{V\left(\Phi _-\right)}}, %-\frac{B \, \sin\chi_+}{2 \sqrt{2} M} k(\Lambda)\sqrt{V\left(\Phi _+\right)}, 
\end{eqnarray}
and an exponential decay/growth term,
\begin{eqnarray}
\Gamma &=& 
%\frac{\sin\chi_+}{4 \sqrt{2} M} \omega (\Lambda) \frac{V'\left(\Phi _+\right)}{\sqrt{V\left(\Phi _+\right)}}+
\frac{B \, k(\Lambda) }{2 \sqrt{2} M}\sqrt{V\left(\Phi _-\right)}.
\end{eqnarray}

The last terms do not contribute if the minimum of the potential, achieved by the field asymptotically, can be set to vanish by a simple shift - it is worth reminding that, even in this situation, the first term remains finite.
Then the $\Gamma$ factor does not affect the time evolution and we only have the oscillating behaviour left.
Since the conserved charges can be obtained as the coefficients in a power series of the spectral parameter $\lambda$ for a function of $a(\lambda)$ only, then one can argue that the charges themselves must also be asymptotically conserved in time.

\begin{figure}[h!]
\begin{center}
\includegraphics[width=0.43\textwidth]{QSGPhiN4.eps}
\includegraphics[width=0.43\textwidth]{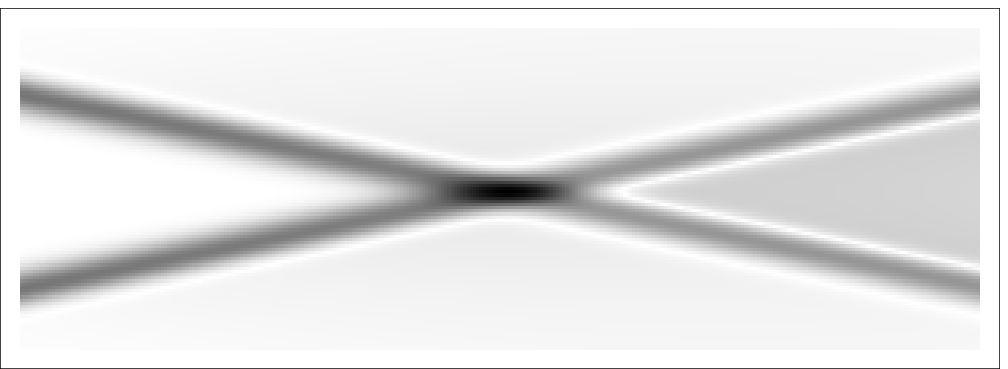}\\
\includegraphics[width=0.43\textwidth]{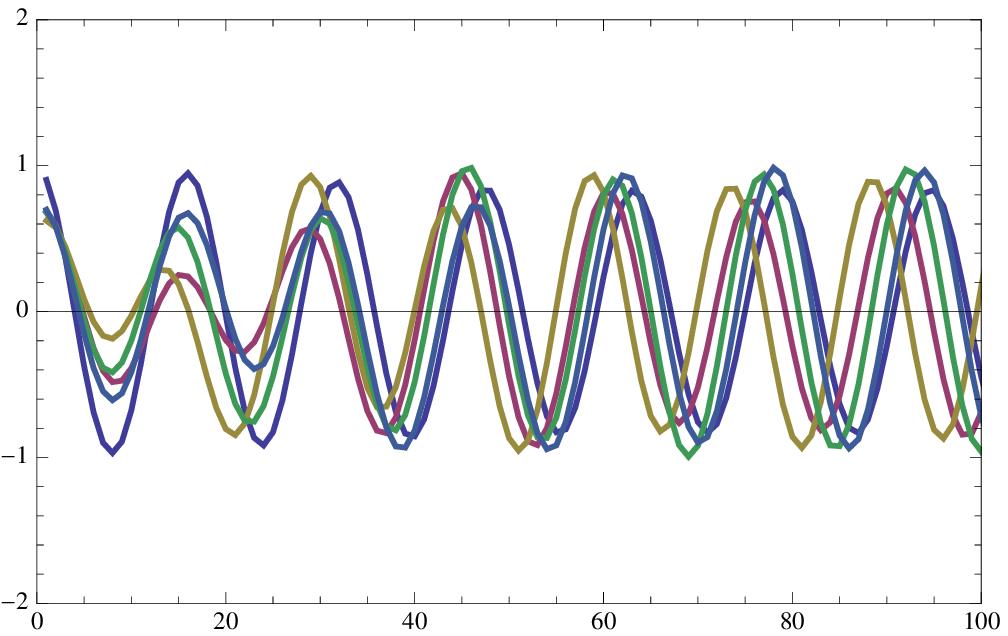}
\includegraphics[width=0.43\textwidth]{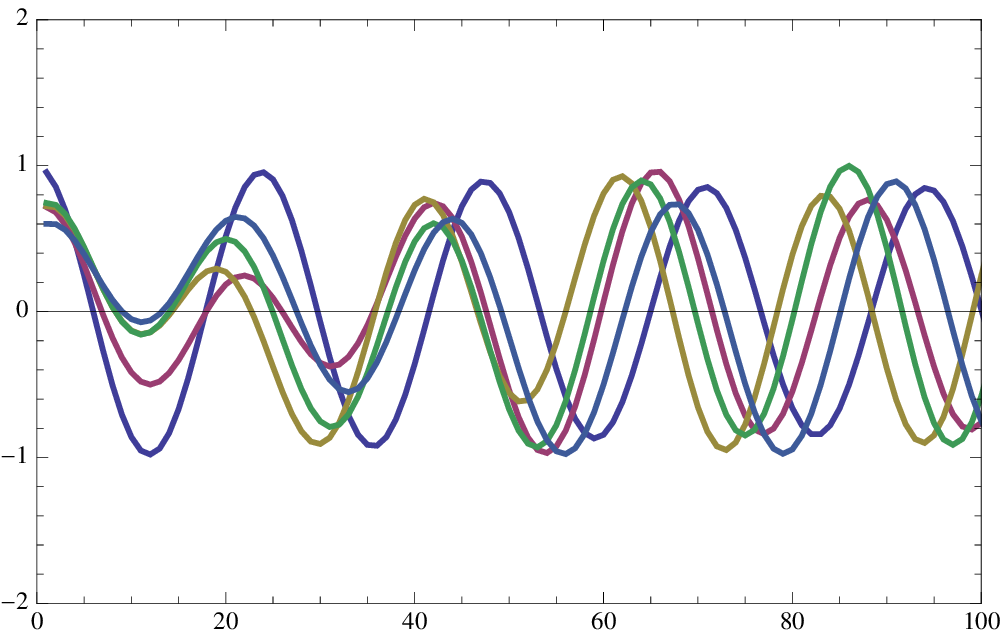}\\
\includegraphics[width=0.43\textwidth]{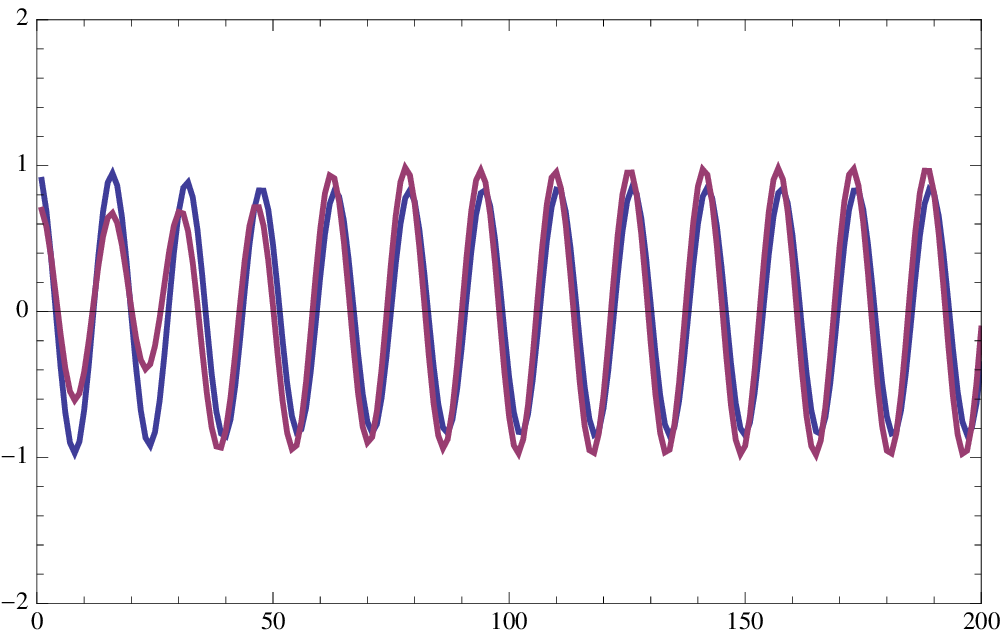}
\includegraphics[width=0.43\textwidth]{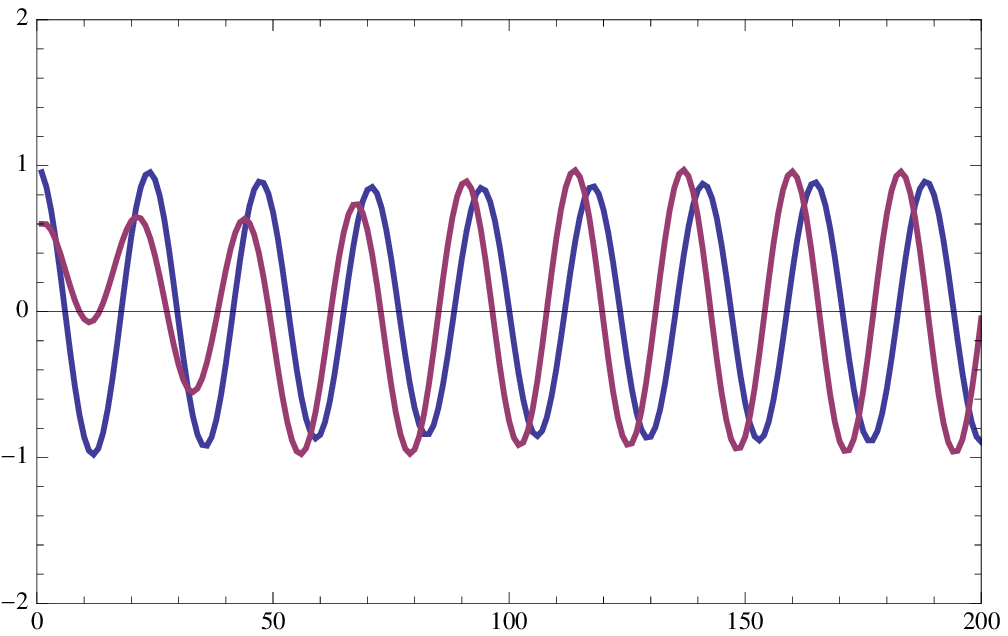}
\caption{On top we show the evolution of an initial configuration under the Bazeia potential for $n=4,6$, respectively.
Then, we depict the spectral curves at five different instants, equally separated, represented by different colours.
We can observe it is not constant in time but one can see after some time the spectral curve tends to recover the initial profile.
Finally, at the bottom the spectral curves at approximately symmetric asymptoitic times.
The greater the deviation from the integrable case (n=2) the more the spectral curves vary.}\label{SpecDatFigure}
\end{center}
\end{figure}

In fact, a simple numerical evaluation shows that, for the model proposed by Bazeia et al, the spectral curve at different instants indicate a small nontrivial time evolution of the scattering data. However, when we compare the asymptotic states the spectral curves tend to coincide, as seen in Figure \ref{SpecDatFigure}.

\section{Numerical validation}\label{numsec}

\qquad In order to verify the results previously stated, here we present some numerical analysis to investigate the scattering of solitary waves which present approximately $\mathcal{PT}$-symmetry.
In Figure \ref{SolWaveScatPlot} we show a density plot of the field excitations as time evolves along the horizontal axis, as colliding wave packets seen from above.
The dynamics is governed by a generalised sine-Gordon interaction of the type found in \cite{Bazeia}.

Having the excitations an approximate $\mathcal{PT}$-symmetry we can check for the $\mathcal{PT}$-invariance on the gauge potentials $A_x(x,t)$ and $A_t(x,t)$.
Therefore, we show in Figures \ref{AxD}, \ref{AxOD}, \ref{AtD}, \ref{AtOD} the behaviour of the diagonal and off-diagonal components of theses connections and confirm the desired properties are observed, at least in a first approximation.

In fact, it becomes evident, however, that as the deformation parameter, $\varepsilon = n-2$, moves away form the integrable situation, $n=2$, the scattering symmetry tends to deteriorate.

In all of these cases, one could also plot the behaviour of the anomaly $\Delta$ coming from the nonflatness of the curvature $F_{\mu \nu}$ and it is possible to verify that its values is relatively small and bounded in this region we are interested.

Finally we present the spectral curve for the model proposed by Bazeia et al \cite{Bazeia}, at two different instants. The figures show that there is indeed a smalll deviation from the conserving regime but the spectral curve, and consequently the conserved charges, tend to recover their assimptotic values.

\section{Concluding remarks}\label{Final}

\qquad The space and time coordinates play a key role in the zero curvature formulation based on the nonabelian Stokes theorem. In this work it was shown that investigating the $\mathcal{PT}$ symmetry of an integrable model at that level might prove interesting as one considers deformations which in principle break the integrability of the system.
We have presented a general scheme and a particular example where asymptotic charges may be considered to be preserved even if the model does not admit a compatible Lax pair. The numerical studies carried out also emphasize the connection between $\mathcal{PT}$-symmetry and the so called \textit{quasi-integrability}.

\vspace{0.5cm}

\noindent {\bf Acknowledgements:}
 The author was partially supported by a Fapesp grant and would like to thank important discussions with L.A.Ferreira, A.Fring and G.Mussardo.
 %P.E.G.A. was partially supported by a Fapesp grant.

% \newpage

\bibliographystyle{unsrt}
\bibliography{reference}

\begin{thebibliography}{99}

\bibitem{BenderBoettcher}
C. M. Bender and S. Boettcher,\textit{Real spectra in non-Hermitian Hamiltonians having $\mathcal{PT}$-symmetry}, Phys. Rev. Lett. \textbf{80} 5243-5246 (1998).
\bibitem{Hook}
C. M. Bender, D. D. Holm, D. Hook, \textit{Complexified dynamical systems}, J. Phys. A: Math. Theor. 40, 2007.
\bibitem{BenderKdV} C. M. Bender et al, \textit{PT-symmetric extension of the Korteweg-de Vries equation}, J. Phys. A: Math. Theor. 40, 2007.
\bibitem{FringKdV}
A. Fring, \textit{PT-symmetric deformations of the Korteweg-de Vries equation}, J. Phys. A: Math. Theor. 40, 2007.
\bibitem{AFKdV}
P. E. G. Assis, A. Fring \textit{Integrable models from PT -symmetric deformations}, J. Phys. A. 42, (2009) 105206 (12pp).
\bibitem{BenderCooper}
C. M. Bender,  F. Cooper, A. Khare, B. Mihaila, A. Saxena, \textit{Compactons in PT-symmetric generalized KdV equations}, Pramana Journal of Physics, 73 (2). pp. 375-385, 2009.
\bibitem{AFCompactons}
P. E. G. Assis, A. Fring, \textit{Compactons versus Solitons}, Pramana, 74(6), pp. 857-865.
\bibitem{AFCalogero}
P. E. G. Assis, A. Fring, \textit{From real fields to complex Calogero particles}, J. Phys. A 42 (2009) 425206 (14pp).
\bibitem{BagchiFringKdV}
B. Bagchi, A. Fring,  \textit{$\mathcal{PT}$-symmetric extensions of the supersymmetric Korteweg-de Vries equation}, J. Phys. A: Math. Theor. 41 392004 (2008). 
\bibitem{CavaFringBagchi} 
A. Cavaglia, A. Fring, B. Bagchi, \textit{$\mathcal{PT}$-symmetry breaking in complex nonlinear wave equations and their deformations}, J. Phys. A: Math. Theor. 44 325201 (2011).
\bibitem{CavagliaFring}
A. Cavaglia, A. Fring, \textit{$\mathcal {PT}$-symmetrically deformed shock waves}, J. Phys. A: Math. Theor. 45 444010 (2012).
\bibitem{PEGA21}
P. E. G. Assis, \textit{A(2|1) spectral equivalences and nonlocal integrals of motion}, J. Phys. A: Math. Theor. 46 195204 (2013).
\bibitem{PTdimer}
J. Pickton, H. Susanto, \textit{Integrability of PT-symmetric dimers}, Phys. Rev. A 88, 063840 (2013).
\bibitem{NonlDynPT}
P. G. Kevrekidis, D. E. Pelinovsky, D. Y. Tyugin, \textit{Nonlinear dynamics in PT-symmetric lattices} J. Phys. A: Math. Theor. 46 365201 (2013).
\bibitem{KinkBreathPT}
D. Saadatmand, S. V. Dmitriev, D. I. Borisov, P. G. Kevrekidis, \textit{Interaction of sine-Gordon kinks and breathers with a parity-time-symmetric defect}, Phys.Rev. E90  052902 (2014).
\bibitem{NonlWang}
Y.Y. Wang, C.Q. Dai, X.G. Wang, \textit{Stable localized spatial solitons in PT -symmetric potentials with power-law nonlinearity},
Nonlinear Dyn 77:1323?1330 (2014).
\bibitem{DisSolVortPT}
Z. Chen, J. Liu, S. Fu, Y. Li, B. A. Malomed, \textit{Discrete solitons and vortices on two-dimensional lattices of PT-symmetric couplers}, Optics Express Vol. 22, Issue 24, pp. 29679-29692 (2014).
\bibitem{PTBreath}
N. Lu, J. Cuevas-Maraver, P. G. Kevrekidis, \textit{PT-symmetric sine-Gordon breathers},  J.Phys. A 47 455101 (2014).
\bibitem{FerrWojZak}
L.A. Ferreira and W.J. Zakrzewski,\textit{The concept of quasi-integrability: a concrete example},JHEP 1105:130 (2011).
\bibitem{Bazeia}
D. Bazeia, L. Losano, J.L.C. Malbouisson and R. Menezes, \textit{Classical behaviour of deformed sine-Gordon models}, Physica D \textbf{237}, 937 (2008).
\bibitem{BoussCal}
H. Airault, H. P. McKean and J. Moser, \textit{Rational and Elliptic Solutions of the KdV Equation and a Related Many-Body Problem}, Comm. Pure Appl. Math. 30, 95-148 (1977).
\bibitem{Alvarez}
O. Alvarez, L.A. Ferreira and J. Sanchez-Guillen, \textit{A new approach to integrable theories in any dimension}, Nuclear Physics B \textbf{529} (3), 689 - 736 (1998).
\bibitem{AssisFerreiraQSG}
P.E.G. Assis and L.A. Ferreira, in preparation.
%\textit{The inverse scattering method for quasi-integrable models},
\bibitem{BabelonBernardTalon}
O. Babelon, D. Bernard and M. Talon, \textit{Introduction to Classical Integrable Systems}, Cambridge Monographs on Mathematical Physics.
\bibitem{Zwanziger}
D. Zwanziger,  \textit{Unstable Particles in S-Matrix Theory}, Phys. Rev. 131, 888 (1963).
\bibitem{Miramontes}
J.L. Miramontes and C. R. Fernández-Pousa, \textit{Integrable quantum field theories with unstable particles}, Physics Letters B 472.3 (2000): 392-401.
\bibitem{OlallaFringUnstable}
O.A. Castro-Alvaredo, Andreas Fring, \textit{Integrable Models with Unstable Particles}, Progress in Mathematics Vol. 237, 2005, pp 59-87.



\end{thebibliography}

\end{document}